%% file: BABAR_PUB_15179.tex
\documentclass[twocolumn,showpacs,aps,prd,floatfix,preprintnumbers]{revtex4}
\usepackage{graphicx} 
\usepackage{dcolumn}
\usepackage{epsfig}    
\usepackage{amsmath}

\input babarsym   % the official way

\newcommand{\psit}{\ensuremath{\psi(2S)}}

\renewcommand{\kk}{\ensuremath{\kp\km\pip\pim\piz}}
\renewcommand{\mm}{\ensuremath{M^2_{\mathrm{miss}}}}
\renewcommand{\ev}{\ensuremath{\mbox{eV}}}

\renewcommand{\gg}{\ensuremath{\gamma\gamma}}
\renewcommand{\X}{\ensuremath{X(3915)}}
\newcommand{\tlp}{\ensuremath{\theta^*_{\ell}}}

\def\ome {\ensuremath{\omega}\xspace}
\def\phil       {\mbox{$\phi_l$}\xspace}
\def\thetal     {\mbox{$\theta_l$}\xspace}
\def\cthetal     {\mbox{$\cos\theta_l$}\xspace}
\def\thetah     {\mbox{$\theta_h$}\xspace}
\def\cthetah     {\mbox{$\cos\theta_h$}\xspace}
\def\thetan     {\mbox{$\theta_n$}\xspace}
\def\cthetan     {\mbox{$\cos\theta_n$}\xspace}
\def\thetalp     {\mbox{$\theta_l^*$}\xspace}

\def\thetanp     {\mbox{$\theta_n^*$}\xspace}
\def\cthetanp     {\mbox{$\cos\theta_n^*$}\xspace}
\def\thetaln     {\mbox{$\theta_{ln}$}\xspace}
\def\cthetaln     {\mbox{$\cos\theta_{ln}$}\xspace}
\def\jome {\ensuremath{\jpsi\omega}\xspace}
\newcommand{\sign}{\ensuremath{7.6}}

\newcommand{\mass}{\ensuremath{3919.4}}
\newcommand{\massStat}{\ensuremath{2.2}}
\newcommand{\massSyst}{\ensuremath{1.6}}
\newcommand{\width}{\ensuremath{13}}
\newcommand{\widthStat}{\ensuremath{6}}
\newcommand{\widthSyst}{\ensuremath{3}}
\newcommand{\ggxvZ}{\ensuremath{52}}
\newcommand{\ggxStatZ}{\ensuremath{10}}
\newcommand{\ggxSystZ}{\ensuremath{3}}
\newcommand{\ggxvT}{\ensuremath{10.5}}
\newcommand{\ggxStatT}{\ensuremath{1.9}}
\newcommand{\ggxSystT}{\ensuremath{0.6}}
\newcommand{\ggul}{\ensuremath{1.7}}

\begin{document}
\newcommand{\BaBarPubYear}    {12}
\newcommand{\BaBarPubNumber}  {001}
\newcommand{\SLACPubNumber}   {15179}

\preprint{BABAR-PUB-\BaBarPubYear/\BaBarPubNumber\\}
\preprint{SLAC-PUB-\SLACPubNumber}

\title{
 \large \bf\boldmath Study of $\X \to \jpsi\omega$ in
 two-photon collisions
}

\input  authors_may2012
\date{September 13, 2012}

\begin{abstract}
We study the process $\gg\to\jpsi\omega$ using a data sample
of 519.2~\invfb\ recorded by the \babar\ detector at SLAC at the PEP-II
asymmetric-energy \epem\ collider at center-of-mass energies near the
$\Upsilon(nS)$ ($n = 2,3,4$) resonances.
We confirm the existence of the charmonium-like resonance \X\ decaying
to $\jpsi\omega$ with a significance of $\sign$
standard deviations, including systematic
  uncertainties, and measure its mass 
$(\mass\pm\massStat\pm\massSyst)~\mevcc$ and width
$(\width\pm\widthStat\pm\widthSyst)~\mev$, where the first uncertainty is 
statistical and the second systematic.
A spin-parity analysis supports the assignment $J^P=0^+$ and therefore the identification of the signal as due to the $\chi_{c0}(2P)$ resonance.
In this hypothesis we determine  the product between the two-photon width and
the final state branching fraction to be 
$(\ggxvZ\pm\ggxStatZ\pm\ggxSystZ)~\ev$ .

\end{abstract}

\maketitle
\section{Introduction}

In the last several years many new charmonium-like states have been observed in the mass
region between 3.7 and 5.0 \gevcc, above
the $D\overline{D}$ threshold, with properties that disfavor their
interpretation as conventional charmonium
mesons~\cite{newStates,BelleZ,BabarZ,Y3940a,Y3940b}. 
The \X\ resonance, decaying to the
$\jpsi\omega$ final state, was first observed by the Belle Collaboration
in two-photon collisions~\cite{BELLE_X3915}.
Another resonance, dubbed $Y(3940)$, has been observed in 
the $B\to\jpsi\omega K$ process~\cite{BABAR_X3872_omega,Y3940a,Y3940b}.
The mass measurement for the $Y(3940)$~\cite{Y3940a,Y3940b,BABAR_X3872_omega} is
consistent with  that of the \X~\cite{BELLE_X3915}. Thus, the same
particle, with a mass of about 
3915~\mevcc, may have been observed in two distinct production
processes. 
The $Z(3930)$ resonance has been discovered in the $\gg\to
D\overline{D}$ process~\cite{BelleZ,BabarZ}. Its interpretation as the
$\chi_{c2}(2P)$, the first radial excitation of the $\phantom{}^3P_2$
charmonium ground state, is commonly accepted~\cite{PDG}.  
Interpretation of the \X\ as the $\chi_{c0}(2P)$~\cite{Liu10} or
$\chi_{c2}(2P)$ state~\cite{Branz11} has been suggested. The latter
implies that the \X\ and $Z(3930)$ are the same particle, observed in
different decay modes. However, the product of the two-photon width
times the decay branching fraction \calB\  for the \X\ reported by
Belle~\cite{BELLE_X3915} is unexpectedly large compared to other
excited \ccbar\ states~\cite{PDG}. 
Interpretation of the \X\ in the framework of molecular models has also
been proposed~\cite{3915Molecule}.

Despite the many measurements available~\cite{PDG}, the nature of the
$X(3872)$ state, which was first observed by Belle~\cite{BelleX},
is still unclear~\cite{Brambilla}.
The observation of its decay
into $\gamma\jpsi$~\cite{X3872_rad} ensures that this particle has positive
$C$-parity. The spin analysis performed by CDF on the decay $X(3872)\to\jpsi\pip\pim$
concludes that only $J^P=1^{+}$ and $J^P=2^{-}$ are consistent with data~\cite{X3872_cdf}.
Similarly, a recent spin analysis performed by Belle~\cite{X3872_belle} concludes that $J^P=1^{+}$ describes
the data as does $J^P=2^{-}$ with one free parameter. 
An analysis of the $\pip\pim\piz$ mass distribution in 
the $X(3872)\to\jpsi\omega$ decay performed by \babar\ favors the
spin-parity assignment $J^P=2^{-}$~\cite{BABAR_X3872_omega}, 
but a $J^P=1^{+}$ spin assignment is not ruled out. 
If $J^P=2^-$, the production of the $X(3872)$ in two-photon collisions
would be allowed.

In this paper we search for the \X\ and $X(3872)$ resonances in the two-photon process 
$\epem\to\epem\gg\to \epem\jpsi\omega$, where $\jpsi \to \ell^+\ell^-, \ (\ell = e\mbox{ or }\mu$) 
and $\omega \to \pi^+ \pi^- \pi^0$.
Two-photon events where the interacting photons are not quasi-real are
strongly suppressed in this analysis by the selection 
criteria described below. This implies that the allowed $J^{PC}$ values of
any produced resonances are $0^{\pm+}$, $2^{\pm+}$, $4^{\pm+}$, ...;
$3^{++}$, $5^{++}$, ...~\cite{Yang}.       
Angular momentum conservation, parity conservation, and charge conjugation
invariance then imply that these quantum numbers also apply to
the final state.

This paper is organized as follows. In Sec. II we give a brief description of the
\babar\ detector. Section III is devoted to the event reconstruction and data selection. In Sec. IV we present the study of the $\jpsi\ome$ system while in Sec. V we perform an angular analysis of \X. The study of systematic uncertainties is described in
Sec. VI. In Sec. VII we summarize the results. 

\section{The \babar\ detector}
The results presented here are based on data collected
with the \babar\ detector %~\cite{BABARNIM}
at the PEP-II asymmetric-energy $e^+e^-$ collider
located at the SLAC National Accelerator Laboratory, and correspond 
to an integrated luminosity of 519.2~\invfb\ recorded at
center-of-mass energies near the $\Upsilon (nS)$ ($n=2,3,4$)
resonances. 
The \babar\ detector is described in detail elsewhere~\cite{BABARNIM}.
Charged particles are detected, and their
momenta are measured, by a five-layer double-sided microstrip detector
and a 40-layer drift chamber, both operating  in the 1.5~T magnetic 
field of a superconducting
solenoid. 
Photons and electrons are identified in a CsI(Tl) crystal
electromagnetic calorimeter (EMC). Charged-particle
identification (PID) is provided by the specific energy loss in
the tracking devices, and by an internally reflecting, ring-imaging
Cherenkov detector. Muons and neutral \KL\ mesons are detected in the
instrumented flux return  of the magnet.
Monte Carlo (MC) simulated events~\cite{geant}, with sample sizes 
more than 10  times larger than the corresponding data samples, are
used to evaluate the signal efficiency and determine background features. 
Two-photon events are simulated  using the GamGam MC
generator~\cite{BabarZ}.

\section{Events Reconstruction and Data Selection}

In this analysis we select events in which the $e^+$ and $e^-$  beam particles are scattered  
at small angles and remain undetected. In the  $\gamma \gamma \to \jpsi\omega$ process,
the \jpsi\ is reconstructed in the $\ell^+\ell^-$ final state, with
$\ell = e\mbox{ or }\mu$, while the $\omega$ is reconstructed in its dominant \pip\pim\piz\ decay
mode. We only consider events where the number of well-measured charged tracks having
a transverse momentum greater than 0.1 \gevc\ is exactly
equal to four.

Neutral pions are reconstructed through the $\piz\to\gamma\gamma$ decay. 
We require the invariant mass of a \piz\ candidate to be in the range
(115--150)~\mevcc, and its energy in the laboratory system to be
larger than 200~\mev. 
The energy in the laboratory frame of the most energetic photon from \piz\
decay is required to be smaller than 1.4~\gev\ in order to suppress
\piz 's not originating from an $\omega$ decay.
We require the energy of the least
energetic photon from \piz\ decay to be in the range
(0.04--0.60)~\gev, and $|\cos\calH_{\piz}|<0.9$, where 
$\calH_{\piz}$ is the angle between the signal \piz\ flight
direction in the laboratory frame and the direction of one of its
daughters in its rest frame.  These requirements 
are optimized by maximizing $S/\sqrt{S+B}$, where $S$ is the number
of MC signal events with a well-reconstructed \piz, and $B$ is the
number of MC signal events where the \piz\ is misreconstructed.
The $\omega$ is reconstructed by
combining two oppositely charged tracks identified as pions with one \piz. The
$\omega$ signal region is defined as $740<m(\pip\pim\piz)<820~\mevcc$.
The \jpsi\ is reconstructed by combining two tracks that are
identified as oppositely charged muons or electrons.
The measured electron energy is corrected to account for energy
deposits in the EMC consistent with bremsstrahlung radiation. We
require the vertex fit probability of the two leptons to be larger
than 0.1\%. The \jpsi\ signal region is defined as $2.95<m(\epem)<3.14$~\gevcc\
for \epem\ and $3.05<m(\mumu)<3.14$~\gevcc\ for \mumu\ events.
An event with a \jome candidate is constructed by fitting the \jpsi\ and $\omega$
candidates to a common vertex. The \piz\ mass is
constrained to its nominal value~\cite{PDG} in this fit. 
Charged particles are required to
originate from the interaction region.
We require the vertex fit probability of the charmonium candidate to
be larger than 0.1\%.

Background arises mainly from random combinations of particles from
\epem\ annihilation, other two-photon collisions, and initial-state
radiation (ISR) processes.  
We discriminate against $\jpsi\pip\pim\piz$ events produced via ISR by  
requiring
$\mm\equiv(p_{\epem}-p_{\mathrm{rec}})^2>2~(\gevcc)^2$, where
$p_{\epem}$ ($p_{\mathrm{rec}}$) is the four-momentum of the initial
state ($\jpsi\omega$ final state).  
We define \pt\ as the transverse momentum, in the \epem\ rest frame, of the \jome candidate with respect to the beam axis.
Well-reconstructed two-photon events are expected to have a low
transverse momentum \pt\ and a small amount of EMC energy $E_{\mathrm{extra}}$,
i.e., energy not associated with the final state particles. We require $\pt<0.2~\gevc$
and $E_{\mathrm{extra}}<0.3~\gev$.
Events originating from residual ISR $\psit\to\jpsi\pip\pim$ decays may create
fake structures in the $\jpsi\omega$ mass spectrum. We therefore remove events in the
mass window $3.675 < m(\jpsi\pip\pim) < 3.700~\gevcc$, where 
$m(\jpsi\pip\pim)=m(\ell^+\ell^-\pip\pim) - m(\ell^+\ell^-) + m(\jpsi)^{\mathrm{PDG}}$ and 
$m(\jpsi)^{\mathrm{PDG}}$ is the nominal \jpsi\ mass~\cite{PDG}.

The $\jpsi\omega$ signal region is defined as the intersection of the
\jpsi\ and $\omega$ signal regions defined above.
In about 10\% of the events we find more than one candidate, and we select the one having the lowest \pt\ value. We obtain 95 events in the $\jpsi\omega$ signal region.

\section{Study of the $\jpsi\omega$ system}

\begin{figure}[!h]
   \begin{center}
     \includegraphics[scale=0.45]{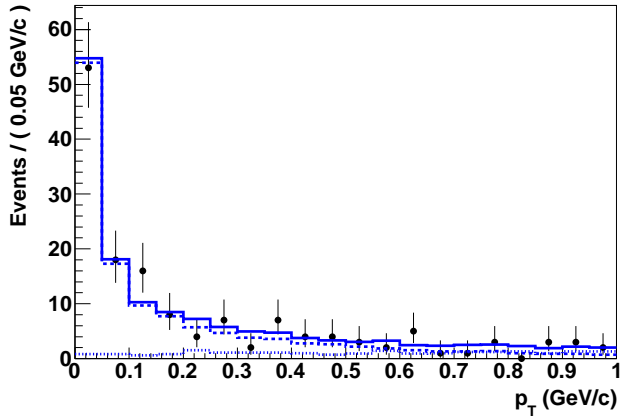}
     \caption{The \pt\ distribution of selected candidates (solid points). 
      The solid histogram represents the result of a fit to
       the sum of the simulated signal (dashed) and background
       (dotted) contributions.} 
   \label{fig:pt}
   \end{center}
 \end{figure}

Figure~\ref{fig:pt}  shows  the \pt\ distribution for the 
selected candidates, obtained by applying the above requirements with
the exception of that on \pt. The
distribution is fitted with the signal \pt\ shape obtained from MC
simulation plus a 
combinatorial background component, modeled using a second-order
polynomial function with free parameters. The number of events from combinatorial
background in the $\pt<0.2~\gevc$ region is $4\pm3$. 

\begin{figure}[!h]
   \begin{center}
\hspace*{-0.5cm}
     \includegraphics[scale=0.45]{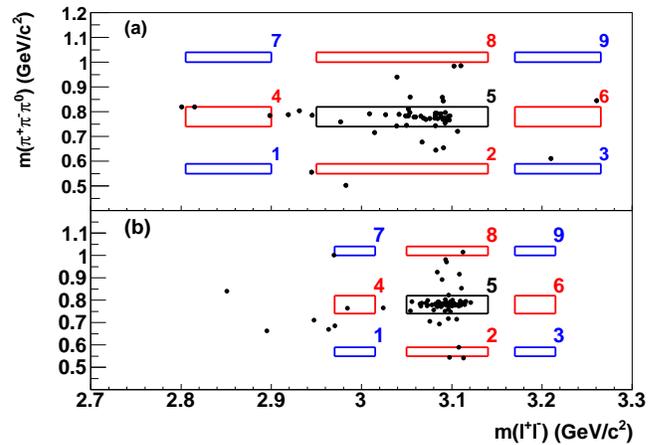}
     \caption{Event distribution (solid points)  in  the $m(\pip\pim\piz)$  versus
       $m(\ell^+\ell^-)$ plane for the (a) \epem and (b)
       \mumu\ decay mode of the  $\jpsi$.
       We also show the $\jpsi\omega$ signal region (tile
       5) and sidebands (tiles 1-4 and 6-9).
     }

   \label{fig:xmass}
   \end{center}
 \end{figure}

\begin{figure}[!h]
   \begin{center}
     \includegraphics[scale=0.45]{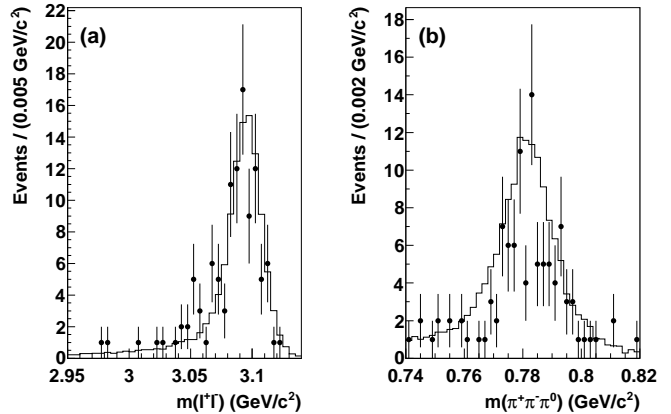}
     \caption{Data (solid points) and normalized MC (histogram) 
       distributions of (a) $m(\ell^+\ell^-)$ for events in the
       $\omega$ signal region, and (b) $m(\pip\pim\piz)$ for events in the
       \jpsi\ signal region. 
     }
   \label{fig:massp}
   \end{center}
 \end{figure}

Figure~\ref{fig:xmass} shows  the
distribution in the 
$m(\ell^+\ell^-)$-$m(\pip\pim\piz)$ plane of events that satisfy the selection
criteria, except for the \jpsi and \ome mass selections. The figure also shows the definitions of signal and background regions, indicated
by the tiles labeled 1-9. The signal regions correspond to tile 5. 
Figures ~\ref{fig:massp}(a) and (b) show
$m(\ell^+\ell^-)$ and $m(\pip\pim\piz)$ for events in the $\omega$ and \jpsi\ signal regions, respectively. 
 As a consistency check, we assign an
$\omega$-Dalitz-plot weight~\cite{BABAR_X3872_omega} to events in the
$\jpsi\omega$ signal region. 
The procedure makes use of the $\omega$
decay angular distribution. The helicity angle $\theta$ is the
angle between the $\pi^+$\ and $\pi^0$\ directions in the $\pi^+\pi^-$
reference frame. The $\cos\theta$ distribution is proportional to
$\sin^2\theta$, and the $\omega$ signal is projected by giving the
$i^{\mathrm{th}}$ event weight $w_i=\frac{5}{2}(1-3\cos^2\theta_i)$. 
The sum of the $\omega$-Dalitz-plot weights is consistent 
with the number of events in the $\jpsi \omega$ signal
region, thus consistent with the hypothesis that most of the observed events do indeed arise  from
true  $\omega \ra \pip\pim\piz$ decays. 


To improve the mass resolution, we define the
reconstructed $\jpsi\omega$ mass as $m(\jpsi\omega) = m(\ell^+\ell^-\pi^+ \pi^- \pi^0)
- m(\ell^+\ell^-) + m(\jpsi)^{\mathrm{PDG}}$. 
The non-$\jpsi\omega$ background is estimated from the \jpsi\ and $\omega$
sidebands defined in Fig.~\ref{fig:xmass}. The $\omega$ sidebands are
defined as [0.55,0.59] \gevcc and [1.00,1.04] \gevcc. The \jpsi sidebands are defined as [2.805,2.900] \gevcc\ and [3.170,3.265] \gevcc\ for the \epem\ channel, [2.970,3.015] \gevcc\ and [3.170,3.215] \gevcc\ for the \mumu\ channel. With these definitions, each sideband size is half of the signal size.  
The $m(\jpsi\omega)$ spectrum of this
background in the $\jpsi\omega$ signal region is obtained by $B(5) =
B(2) + B(4) + 
B(6) + B(8) - (B(1) + B(3) + B(7) + B(9))$, where $B(i)$ is the $m(\jpsi\omega)$
spectrum in the $i^{\mathrm{th}}$ region shown in fig.~\ref{fig:xmass}. The estimated background from this method is
$5\pm 3$ in good agreement with the estimate from the fit to the $p_T$ distribution.
The residual background from
$\psit\to\jpsi\pip\pim$ decay is estimated by using the values of the
integrated luminosity, MC efficiencies, the cross section for \psit\
production in ISR events~\cite{psitXsec}, and the nominal branching fractions
 for the
relevant \psit\ and \jpsi\ decays~\cite{PDG}. The expected number of
background events from such process is smaller than 0.9 at 90\%
confidence level (CL). 

The detection efficiency depends on $m(\jpsi\omega)$ and \tlp, where \tlp\ is the angle between
the direction of the positively charged lepton from \jpsi\
decay ($\ell^+$) and the beam axis in the $\jpsi\omega$ rest frame.
Since we select events in which the $e^+$ and $e^-$  beam particles are scattered  
at small angles, the two-photon axis is
approximately the same as the beam  axis. Therefore we use  the beam axis to determine  \tlp.

We parameterize the efficiency dependence with a two-dimensional ($m(\jpsi\omega)$, \tlp)
histogram. We label MC events where the reconstructed decay particles
are successfully matched to the generated ones as truth-matched events.
The detection efficiency in each histogram bin is
defined as the ratio between the number of truth-matched MC events that satisfy the selection criteria
and the number of MC events that were generated for that bin.

\begin{figure}[!h]
   \begin{center}
     \includegraphics[scale=0.45]{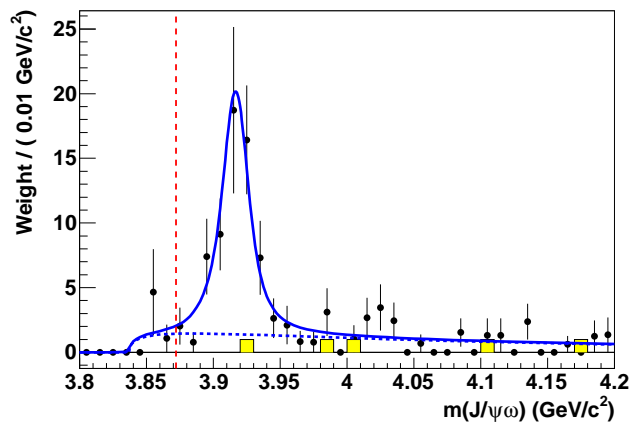}
     \caption{The efficiency-corrected $m(\jpsi\omega)$ distribution of selected  
       events (solid points). The solid line represents
       the total fit function. The dashed line is the NR
       contribution. The shaded histogram is the non-$\jpsi\omega$ background
       defined in the text as $B(5)$ and
       estimated from sidebands. The vertical dashed (red) line is
       placed at $m(\jpsi\omega)=3.872~\gevcc$.}
   \label{fig:fit}
   \end{center}
 \end{figure}

The $m(\jpsi\omega)$ spectrum is shown in Fig.~\ref{fig:fit}, where each event is
weighted to account for  detector efficiency, which 
is almost uniform as a function of the $\jpsi\omega$ mass.
The event weight is equal to $\overline{\eps}/\eps(m(\jpsi\omega),\tlp)$,
where $\eps(m(\jpsi\omega),\tlp)$ is the $m(\jpsi\omega)$- and \tlp-dependent efficiency value 
and $\overline{\eps}$ is a common scaling factor that ensures all the
weights are ${\cal O}(1)$, since weights far from one 
can cause the estimate of the statistical uncertainty to be 
incorrect~\cite{frodesen}. 
We observe a prominent peak near $3915~\mevcc$ over a small
background. No evident structure is observed around
$3872~\mevcc$. 

We perform an extended unbinned maximum-likelihood fit to the
efficiency-corrected $m(\jpsi\omega)$ spectrum to extract the resonance yield and
parameters. In the likelihood function $\calL$ there are two components:
one for the $X(3915)$ signal and one for the non-resonant $\jpsi\omega$ contribution (NR). 
The probability density function (PDF) for the signal component is defined by the convolution
of an $S$-wave relativistic Breit-Wigner distribution 
with a detector resolution function. The NR contribution is
taken to be proportional to $\calP_{bg}(m) = p^*(m)\times \exp[-\delta p^*(m)]$, where
$p^*(m)$ is the \jpsi\ momentum in the rest frame of a $\jpsi\omega$ system with
an invariant mass $m$,
 $\delta$ is a fit parameter, and $m=m(\jpsi\omega)$. The signal and NR
yields, the $X(3915)$ mass and width, and $\delta$ are free parameters in the fit.

We use truth-matched MC events to determine the signal PDF detector
resolution function. The signal detector-resolution PDF is described by 
  the sum of two Gaussian shapes for the $X(3915)$ and the sum of a Gaussian
plus a Crystal Ball function~\cite{CBShape} for the $X(3872)$.
The parameters of the resolution 
  functions are determined from fits to truth-matched MC events.
The widths of the Gaussian core components are
5.7~\mev\ and  4.5~\mev, respectively, for $X(3915)$ and $X(3872)$. 
No significant difference in the resolution function
parameters is observed for the different \jpsi\ decay modes. 
The parameters of the resolution functions are fixed to their
MC values in the maximum-likelihood fit. 

The fitted distribution from the maximum-likelihood fit to the efficiency-corrected $m(\jpsi\omega)$ spectrum 
is shown in Fig.~\ref{fig:fit}.  
We observe $59\pm10$
signal events; the measured \X\ mass and width are
$(3919.4\pm2.2)~\mevcc$ and $(13\pm6)~\mev$, respectively,  where the uncertainties
are statistical only. 
We add an $X(3872)$ component,
modeled as a $P$-wave relativistic Breit-Wigner with
mass $3872~\mevcc$ and width $2~\mev$~\cite{PDG}, convoluted with
the detector resolution function. No significant
change in the result is observed with the addition
of this component, whose yield is estimated to be $1 \pm 4$ events.
An excess of events over the fitted NR is
observed at $m(\jpsi\omega)\sim4025~\mevcc$. 
If we add a resonant component in the likelihood function to fit this
excess, modeled as a Gaussian having free parameters,
we obtain a signal yield of $5\pm3$
events. 

\section{Angular Analysis of the {\bf X(3915)}}

\begin{figure}[!h]
   \begin{center}
     \includegraphics[scale=0.45]{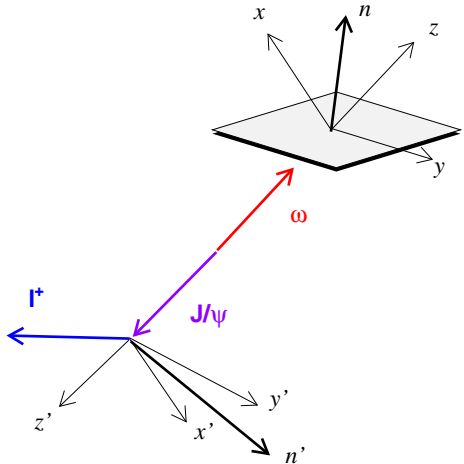}
     \caption{Diagram illustrating the reference frames involved in the definition of angular variables.}
   \label{fig:phi_l}
   \end{center}
 \end{figure}

\begin{figure}[!h]
   \begin{center}
     \includegraphics[scale=0.45]{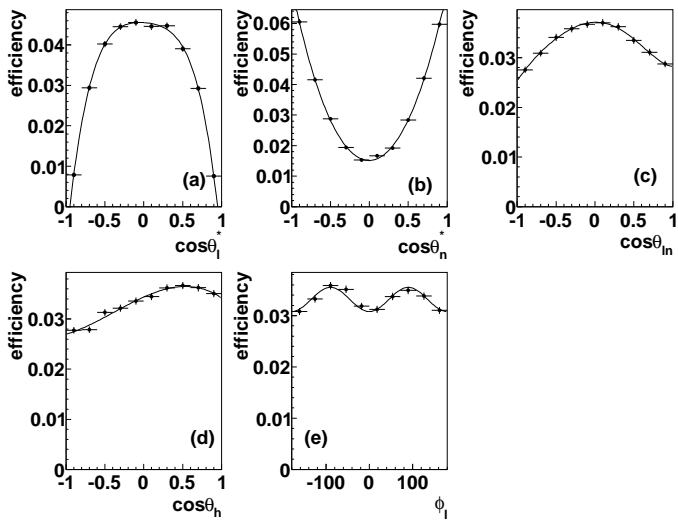}
     \caption{The efficiency distributions in the \X\ signal region $3890<m(\jpsi\omega)<3950~\mevcc$ (solid points) as functions of:
(a) cos\tlp\, (b) \cthetanp, (c) \cthetaln, (d) \cthetah, and (c) \phil. The curves show the results from the fits described in the text. 
     }
   \label{fig:effth}
   \end{center}
 \end{figure}

We first attempt to discriminate between $J^P=0^{\pm}$ and $J^P=2^+$ by using the Rosner~\cite{Rosner} 
predictions. In addition to the previously defined \tlp\, we consider
the following two angles: \thetanp defined as the angle between the normal to the decay plane of the \ome ($\vec{n}$) and the two-photon axis, and \thetaln defined as the angle between the lepton $\ell^+$ from \jpsi decay and the \ome decay normal (see Fig.~\ref{fig:phi_l}).
To obtain the normal to the $\omega$ decay plane we boost the two pions from the \ome decay 
into the $\omega$ rest frame and obtain $\vec{n}$ by the cross product vector of the two charged pions.
A projection of the efficiency values over cos\tlp\ in the \X\ signal region is shown in
Fig.~\ref{fig:effth}(a). The projections of the efficiency
over the angles \thetanp and  \thetaln are shown in Figs.~\ref{fig:effth}(b) and Fig.~\ref{fig:effth}(c).
The efficiency distributions are not uniform and are parameterized by fifth-order polynomials.
 \begin{figure}[!h]
   \begin{center}
     \includegraphics[scale=0.45]{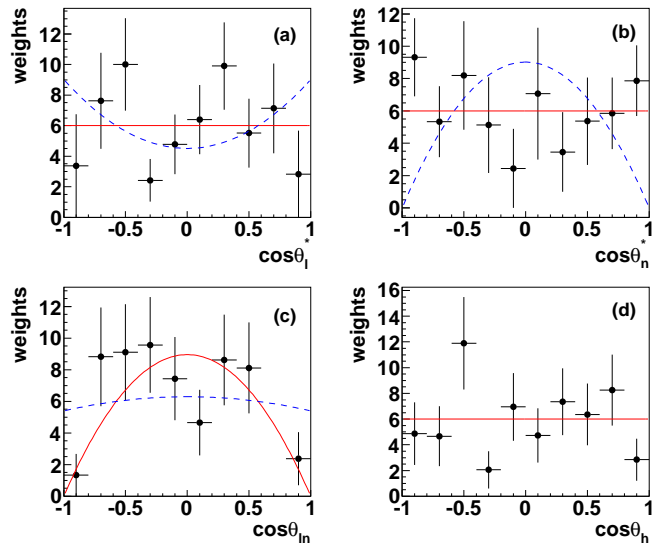}
     \caption{The efficiency-corrected distributions of selected 
       events in the \X \ signal region $3890<m(\jpsi\omega)<3950~\mevcc$ (solid points). (a) cos\tlp\, (b) \cthetanp, (c) \cthetaln, and (d) \cthetah. 
 The solid (red) line represents the
       expected distribution for the $J^P=0^{\pm}$ assignment and the
       dashed (blue) line for the $J^P=2^+$.}
   \label{fig:ang1}
   \end{center}
 \end{figure}
The cos\tlp\, \cthetanp, and \cthetaln distributions are sensitive to the spin-parity of the
resonance. We assume that for $J^P=2^+$ the dominant amplitude has
helicity 2. This is in agreement with previous charmonium measurements~\cite{Ue08b,AubertDD,Ablikim}, and theoretical predictions~\cite{Poppe,Schuler}. The expected functional forms under this hypothesis are
summarized in Table~\ref{tab:spin}.
Figures \ref{fig:ang1}(a),(b), and (c) show the efficiency-corrected cos\tlp\, \cthetanp, and \cthetaln
distributions for events in the \X\ signal region, defined by $3890<m(\jpsi\omega)<3950~\mevcc$. 
Since the background is small, we assume 
that all the events come from \X\ decay.
The distributions for data are compared with
the expected curves for $J^P=0^{\pm}$ and $J^P=2^+$.
The resulting $\chi^2$ for each distribution is reported 
in Table~\ref{tab:spin}. In all cases the $J^P=0^{\pm}$ expectations describe the data better than the $J^P=2^+$
ones and this is particularly true for the \cthetanp distribution. In the latter case $\chi^2$ probabilities for $J^P=0^{\pm}$ and $J^P=2^+$
are respectively 64.7\% and $9.6\times 10^{-9}$\% respectively. 
We conclude that the data largely prefer $J^P=0^{\pm}$ over $J^P=2^+$.
\begin{table}
\caption{Functional shapes and $\chi^2$ for the different spin hypotheses. NDF=9.}
\label{tab:spin}
\begin{center}
\begin{tabular}{lcccc}
\hline \noalign{\vskip2pt}
Angle & $J^P=0^-$ & $J^P=0^+$ & $J^P=0^{\pm}$ & $J^P=2^+$\cr
\hline
\thetalp & & & 1 &  $1+\cos^2\thetalp$\cr
$\chi^2$ & & & 11.2 & 16.9 \cr
\hline
\thetanp & & & 1 &  $\sin^2\thetanp$\cr
$\chi^2$ & & & 6.9 & 65.9 \cr
\hline
\thetaln & & & $\sin^2\thetaln$\ &  $7 - \cos^2\thetaln$\cr
$\chi^2$ & & & 12.5 & 18.0 \cr
\hline
\thetah &  &  & 1 & \cr
$\chi^2$ &  &  & 12.2 & \cr
\hline
\thetan & $\sin^2\thetan$ & 1 \cr
$\chi^2$ & 77.6 & 16.3 & & \cr
\hline
\thetal & $1 + \cos^2\thetal$  & 1 \cr
$\chi^2$ & 8.7 & 8.3& & \cr
\hline
\phil & $2 - \cos(2\cos\phil)$ & $2 + \cos(2\cos\phil)$\cr
$\chi^2$ & 21.7 & 9.6 & &\cr
\hline
\end{tabular}
\end{center}
\end{table}

%%%%%%%%%%%%%%%%%%%%%%%%%%%%%%%%%%%%%
% syst
%%%%%%%%%%%%%%%%%%%%%%%%%%%%%%%%%%%%%

The spin-0 hypothesis can be further tested by examining the \cthetah distribution, where \thetah is the angle formed by the \jpsi momentum in the \jome rest frame with respect to the \jome direction in the laboratory frame. The efficiency
distribution as a function of \cthetah is shown in Fig.~\ref{fig:effth}(d),
where it is parameterized by a third-order polynomial.
The \cthetah distribution in the \X\ signal region, corrected for efficiency, is shown in fig.~\ref{fig:ang1}(d) and is compared with the uniform distribution expected for the spin-0 hypothesis. 
The resulting $\chi^2/NDF$ is 12.2/9 and we conclude that this test also supports the spin-0 assignment.

We attempt to discriminate between $J^P=0^-$ and $J^P=0^+$. For this purpose, we define the angles, \thetan, \thetal, and \phil. To define these angles,
we first boost all the 4-vectors into the \jome rest frame. We define $\theta_n$ to be the 
angle between the normal to the \ome decay plane $\vec{n}$ and the $\omega$ direction in the 
$J/\psi \omega$ rest frame. The efficiency distribution as a function of \cthetan (not shown) is consistent with being uniform.

For \jpsi decay, we first boost the $\ell^+$ to the \jpsi rest frame. We define \thetal as the angle
between the $\ell^+$ and the direction of the \jpsi in the \jome frame.
The efficiency distribution as a function of \cthetal (not shown) is consistent with being uniform.

Next we define a coordinate system as follows (see Fig.~\ref{fig:phi_l}).
For \ome decay, we choose the $z$ axis along the \ome momentum vector,
and represent the \ome decay in terms of its decay plane normal $\vec n$.  
The cross product $\vec{z}\times\vec{n}$ gives the $y$-axis direction.
Then we define the $x$-axis vector by $\vec{y} \times \vec{z}$.
The  $x-z$ plane, by construction, contains the
\ome decay plane normal.

We now specify the \jpsi decay coordinate system in terms of the unit vectors
defined for \ome decay.
We define $\vec z'=-\vec z$, $\vec x'= -\vec x$, and $\vec y'=\vec y$ so that $\vec y'$ is along the normal
to the plane containing the normal to the decay plane of the \ome.
Next we define the \jpsi decay plane normal $\vec n'$ as the cross product of the $\ell^+$
in the \jpsi rest frame and the $\vec z'$ vector. By construction, $\vec n'$ is in the $x'-y'$ plane. Then we compute the angle \phil as the angle between the \jpsi and \ome
decay plane normals.

The efficiency distribution as a function of \phil is shown in Fig.~\ref{fig:effth}(e) and is fitted
using the function $\epsilon(\phi_l)=1-c \cdot \cos2\phi_l$, where $c$ is a free parameter.

It can be shown that the full angular distribution for $J^P=0^-$ can be written as:
\begin{equation}
\frac{dN}{d\cthetan d\cthetal d\phil}=\frac{9N}{64\pi}\sin^2\thetan[1+\cthetal^2 + \sin^2\thetal\cdot \cos2\phil].
\end{equation}

\begin{figure}
   \begin{center}
     \includegraphics[scale=0.45]{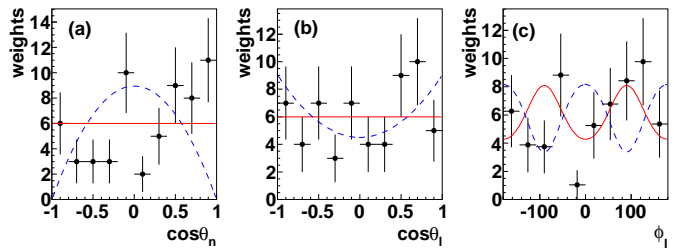}
     \caption{The efficiency-corrected distributions of selected 
       events in the \X\ signal region $3890<m(\jpsi\omega)<3950~\mevcc$ (solid points). 
The (a) \cthetan, (b) \cthetal, and (b) \phil distributions are compared with $J^P=0^+$ (solid line, red) and $J^P=0^-$ (dashed line, blue) expectations.}
   \label{fig:ang2}
   \end{center}
 \end{figure}

For $J^P = 0^+$, assuming no $D$-wave, the normalized angular distribution is given by:
\begin{equation}
\begin{aligned}
\frac{dN}{d\cthetan d\cthetal d\phil}=&\frac{3N}{32\pi}[2\sin^2\thetal \cos^2\thetan + \\
 &\sin^2 \thetan \cdot(1+\cos^2\thetal - \sin^2\thetal \cos 2\phil) + \\ 
 & \sin 2\thetal \cos\phil \sin 2\thetan].
\end{aligned}
\end{equation}
Eqs. (1) and (2), when projected onto the different angles, give the functional expectations shown in Table~\ref{tab:spin} and presented 
in Fig.~\ref{fig:ang2}. The resulting $\chi^2$ for all the distributions are summarized in Table~\ref{tab:spin}. In all cases the $J^P=0^+$ hypothesis gives a smaller
$\chi^2$ than the $J^P=0^-$ hypothesis and this is particularly true for the \cthetan distribution. 
In the latter case $\chi^2$ probabilities for $J^P=0^+$ and $J^P=0^-$
are 6.1\% and 4.8$\times10^{-11}$\% respectively. 
We conclude that the $J^P=0^+$ assignment is largely preferred over the $J^P=0^-$
assignment.

We observe no correlation between any angles considered in this analysis except for \phil which is strongly correlated with \thetaln.

\section{Systematic Uncertainties}

Several sources contribute to systematic uncertainties on the
resonance yields and parameters.
Systematic uncertainties due to the functional forms chosen for the PDF parametrizations and fixed
parameters in the fit are estimated to be the sum in quadrature of the
changes observed when repeating the fit varying the
fixed parameters by $\pm1$ standard deviation ($\sigma$). 
Since the \X\ spin assignment is unknown, we repeat the fit by
parameterizing the \X\ signal as the convolution of a $P$-wave
relativistic Breit-Wigner with the detector resolution
function. The changes in the fit results are taken as the systematic uncertainty.
We examine the dependence of the fit results on the fit range, varying the boundary of the fit
from the nominal value of
$4.2~\gevcc$ (see Fig.~\ref{fig:fit}) to either $4.1$ or $4.3~\gevcc$.
We take as the systematic uncertainty the largest among the observed
differences in the fit results. 
The uncertainty on the absolute mass scale is 
studied by measuring the difference between the observed and nominal
\jpsi\ mass in a \kk\ ISR-enriched control sample~\cite{prevWork}. The 
\kk\ final state has the same number of charged and neutral
particles as $\jpsi\omega$. The observed difference in mass is 
$(-1.1\pm0.8)~\mevcc$. We take the sum in quadrature of this shift
with its uncertainty as a systematic uncertainty.
Previous studies show that MC events have a better
mass resolution than data~\cite{prevWork}. 
The effect of possible data/MC differences in the $m(\jpsi\omega)$ resolution is
estimated by increasing the width of the resolution function
core component by 20\%. 
The uncertainty due to the use of efficiency weights to correct the
$m(\jpsi\omega)$ spectrum is estimated with simulated experiments. 
In each experiment, we randomly modify the efficiency weight according to its statistical uncertainty. 
We then fit the resulting mass spectra and plot the resulting yields and resonance parameters.
The resulting spreads give the systematic uncertainties on these quantities. 
We find that the fit bias on the yield is negligible. 

The \X\ signal significance is $\sign\sigma$, calculated from  
$-2\ln{(\calL_{0}/{\calL}_{\max})}$, where $\calL_{0}$ and $\calL$ are the likelihoods
of the fits with and without the resonant component, respectively.
The difference in the number of degrees of freedom is taken into account.
Systematic uncertainties are incorporated into the likelihood function by convolving it with a Gaussian
with mean equal to zero and width equal to the systematic uncertainty on the yield.

The product between the two-photon coupling $\Gamma_{\gg}$ and the
resonance branching fraction $\calB$ to the $\jpsi\omega$ final state
is measured  using $473.8~\invfb$ of data collected near the \FourS\
energy.  
The efficiency-weighted yields for the resonances, the integrated
luminosity near the \FourS\ energy, and the branching fractions 
$\calB(\jpsi\to\ell^+\ell^-)=(5.94\pm0.06)\%$~\cite{PDG} and
$\calB(\omega\to\pip\pim\piz)=(89.2\pm0.7)\%$~\cite{PDG} 
are used to obtain $\Gamma_{\gg}\times\calB$ using the GamGam
generator. In this calculation, the \X\ parameters
are fixed to the values obtained from the fit.

The uncertainties on the weighted signal yield described
above are taken into account in the $\Gamma_{\gg}\times\calB$
systematic error. Systematic uncertainties on the efficiency due to
tracking (0.3\%  per track), \piz\ reconstruction
(3.0\%) and PID (0.1\% per pion, 0.8\% per lepton) are obtained from
auxiliary studies. 
The uncertainty on the luminosity is 1.1\%. The uncertainty on the
nominal \jpsi\ and $\omega$ branching fractions used in the calculation are
propagated in the 
$\Gamma_{\gg}\times\calB$ error.
The GamGam calculation has an uncertainty of 3\%~\cite{BabarZ}.

Since no significant $X(3872)$ signal is observed, we determine a
Bayesian upper limit (UL) at 90\% CL on
$\Gamma_{\gg}\times\calB$, assuming a uniform prior probability
distribution. 
The upper limit for $\Gamma_{\gg}\times\calB$ is thus computed
according to:
$$\int_0^{\rm UL} L(\Gamma_{\gg}\times\calB) d(\Gamma_{\gg}\times\calB) =0.90$$
where $L(\Gamma_{\gg}\times\calB)$ is the likelihood function for
$\Gamma_{\gg}\times\calB$.

For a $J = 0$ resonance, the resulting value of $\Gamma_{\gg}[\X]\times\calB(\X\to\jpsi\omega)$
is $(\ggxvZ\pm\ggxStatZ\pm\ggxSystZ)~\ev$ where the first uncertainty is statistical and the second systematic. For completeness we also report the value for $J = 2$:   
$(\ggxvT\pm\ggxStatT\pm\ggxSystT)~\ev$.
For $X(3872)$,  we obtain 
$\Gamma_{\gg}[X(3872)]\times\calB(X(3872)\to\jpsi\omega) < \ggul~\ev$ at 90\% CL,
assuming $J=2$.

\section{Summary}

In summary, we confirm the observation of the
charmonium-like resonance \X\ in the $\gg\to\jpsi\omega$ process, with
a significance of $\sign\sigma$, including systematic
  uncertainties. The measured mass and width are
\begin{eqnarray}
m[\X]&=&(\mass\pm\massStat\pm\massSyst)~\mevcc, \nonumber\\
\Gamma[\X]&=&(\width\pm\widthStat\pm\widthSyst)~\mev, \nonumber 
\end{eqnarray}
where the first uncertainty is statistical and the second systematic. These
measurements are consistent with those previously reported by
Belle for the same process~\cite{BELLE_X3915} and by \babar\ ~\cite{Y3940b} and Belle~\cite{Y3940a} for $B \to \jpsi \ome K$.
A detailed angular analysis has been performed. 
We find that the data largely prefer $J^P=0^{\pm}$ over $J^P=2^+$. 
In this hypothesis, $J^P=0^+$ is largely preferred over $J^P=0^-$ and this would 
identify the signal as being due to the  $\chi_{c0}(2P)$ resonance.
The mass of \X\ is consistent
with the result of the potential model, which predicts
the mass of the first radial excitation $\chi_{c0}$ to be around 3916 MeV
according to Godfrey-Isgur relativized potential model~\cite{Isgur}.
 The product 
$\Gamma_{\gg}[\X]\times\calB[\X\to\jpsi\omega]$ is also measured. The
value for $J=0$ (relatively large compared 
to charmonium model predictions)
is consistent with that reported by Belle~\cite{BELLE_X3915}. This product,
also computed in this analysis for $J=2$, is smaller than the corresponding value obtained by Belle.
We have also searched for the $\gg\to X(3872)\to\jpsi\omega$
process, but no significant signal is found.

\section{Acknowledgements}
We are grateful for the 
extraordinary contributions of our \pep2\ colleagues in
achieving the excellent luminosity and machine conditions
that have made this work possible.
The success of this project also relies critically on the 
expertise and dedication of the computing organizations that 
support \babar.
The collaborating institutions wish to thank 
SLAC for its support and the kind hospitality extended to them. 
This work is supported by the
US Department of Energy
and National Science Foundation, the
Natural Sciences and Engineering Research Council (Canada),
the Commissariat \`a l'Energie Atomique and
Institut National de Physique Nucl\'eaire et de Physique des Particules
(France), the
Bundesministerium f\"ur Bildung und Forschung and
Deutsche Forschungsgemeinschaft
(Germany), the
Istituto Nazionale di Fisica Nucleare (Italy),
the Foundation for Fundamental Research on Matter (The Netherlands),
the Research Council of Norway, the
Ministry of Education and Science of the Russian Federation, 
Ministerio de Ciencia e Innovaci\'on (Spain), and the
Science and Technology Facilities Council (United Kingdom).
Individuals have received support from 
the Marie-Curie IEF program (European Union), the A. P. Sloan Foundation (USA) 
and the Binational Science Foundation (USA-Israel).

\renewcommand{\baselinestretch}{1}

\end{document}

%% file: authors_may2012.tex
%% author list as of 02-May-2012 (363 authors)
%
\author{J.~P.~Lees}
\author{V.~Poireau}
\author{V.~Tisserand}
\affiliation{Laboratoire d'Annecy-le-Vieux de Physique des Particules (LAPP), Universit\'e de Savoie, CNRS/IN2P3,  F-74941 Annecy-Le-Vieux, France}
\author{J.~Garra~Tico}
\author{E.~Grauges}
\affiliation{Universitat de Barcelona, Facultat de Fisica, Departament ECM, E-08028 Barcelona, Spain }
\author{A.~Palano$^{ab}$ }
\affiliation{INFN Sezione di Bari$^{a}$; Dipartimento di Fisica, Universit\`a di Bari$^{b}$, I-70126 Bari, Italy }
\author{G.~Eigen}
\author{B.~Stugu}
\affiliation{University of Bergen, Institute of Physics, N-5007 Bergen, Norway }
\author{D.~N.~Brown}
\author{L.~T.~Kerth}
\author{Yu.~G.~Kolomensky}
\author{G.~Lynch}
\affiliation{Lawrence Berkeley National Laboratory and University of California, Berkeley, California 94720, USA }
% ...
\collaboration{The \babar\ Collaboration}
\noaffiliation

%% file: BABAR_PUB_15179.bbl
\begin{thebibliography}{99}

\bibitem{newStates}
%Y(4260)
B.~Aubert \etal\ (\babar\ Collaboration), \jprl{95}, 142001 (2005);
T.~E.~Coan \etal\ (CLEO Collaboration), \jprl{96}, 162003 (2006);
C.~Z.~Yuan \etal\ (Belle Collaboration), \jprl{99}, 182004 (2007);
%Y(4350)
B.~Aubert \etal\ (\babar\ Collaboration), \jprl{98}, 212001 (2007);
X.~L.~Wang \etal\ (Belle Collaboration),  \jprl{99}, 142002 (2007).

\bibitem{BelleZ}
S.~Uehara \etal\ (Belle Collaboration), \jprl{96}, 082003 (2006).

\bibitem{BabarZ}
B.~Aubert \etal\ (\babar\ Collaboration), \jprd{81}, 092003 (2010).

\bibitem{Y3940a}
S.-K.~Choi \etal\  (Belle Collaboration),  \jprl{94}, 182002 (2005);
\bibitem{Y3940b}
B.~Aubert \etal\ (\babar\ Collaboration), \jprl{101}, 082001 (2008).

\bibitem{BELLE_X3915}
S.~Uehara \etal\ (Belle Collaboration), \jprl{104}, 092001 (2010).

\bibitem{BABAR_X3872_omega}
P.~del~Amo~Sanchez \etal\ (\babar\ Collaboration), \jprd{82}, 011101(R)
(2010).
\bibitem{PDG}
Particle Data Group, K. Nakamura \etal\, J. Phys. G {\bf 37}, 075021
(2010).  
\bibitem{Liu10} %3915 is chi_c0(2P)
X.~Liu \etal\, \jprl{104}, 122001 (2010).

\bibitem{Branz11}%3915 is chi_c2(2P)
T.~Branz \etal\, \jprd{83}, 114015 (2011).



\bibitem{3915Molecule}
X. Liu \etal\, \epjc{61}, 411 (2009);
T.~Branz \etal\, \jprd{80}, 054019 (2009);
W.~H.~Liang \etal\, Eur.~Phys.~Jour.~{\bf A 44}, 479 (2010).


\bibitem{BelleX}
S.-K.~Choi \etal\ (Belle Collaboration), \jprl{91}, 262001 (2003).

\bibitem{Brambilla}
N.~Brambilla \etal\, \epjc{71}, 1534 (2011).

\bibitem{X3872_rad}
B.~Aubert \etal\ (\babar\ Collaboration), \jprl{102}, 132001
(2009);
V.~Bhardwaj \etal\ (Belle Collaboration), \jprl{107}, 091803 (2011). 

\bibitem{X3872_cdf}
A.~Abulencia \etal\ (CDF Collaboration), \jprl{98}, 132002 (2007).

\bibitem{X3872_belle}
S.-K.~Choi \etal\  (Belle Collaboration),  \jprd{84}, 052004 (2011).

\bibitem{Yang}
C.~N.~Yang, \pr{77}, 242 (1950).

\bibitem{BABARNIM}
B.\ Aubert \etal\ (\babar\ Collaboration), \nima{479}, 1 (2002).

\bibitem{geant}
The \babar\ detector Monte Carlo simulation is based on GEANT4
[S. Agostinelli \etal, \nima{506}, 250 (2003)] and EvtGen [D.~J.~Lange,
\nima{462}, 152 (2001)].

\bibitem{psitXsec}
J.~P.~Alexander \etal, SLAC-PUB-4501 (1998);
M.~Benayoun \etal, Mod.~Phys.~Lett.~{\bf A14}, 2605 (1999).

\bibitem{frodesen}
A.~G.~Frodesen \etal, \emph{Probability and Statistics in Particle
  Physics} (Universitetsforlaget, Bergen, Norway, 1979).

\bibitem{CBShape}
M.~J.~Oreglia, Ph.D Thesis, SLAC-236(1980), Appendix D;
J.~E.~Gaiser, Ph.D Thesis, SLAC-255(1982), Appendix F;
T.~Skwarnicki, Ph.D Thesis, DESY F31-86-02(1986), Appendix E.

\bibitem{Rosner}
J.~L.~Rosner, \jprd{70}, 094023 (2004).

\bibitem{Ue08b} S.~Uehara \etal\ (Belle Collaboration), \jprd{78}, 052004 (2008).
\bibitem{AubertDD} B.~Aubert \etal\ (\babar\ Collaboration), \jprd{81}, 092003 (2010).
\bibitem{Ablikim} M.~Ablikim \etal\ (BESIII Collaboration), arXiv:1205.4284v1 (2012).
\bibitem{Poppe} M.~Poppe, \ijmpa{1}, 545 (1986).
\bibitem{Schuler} G.A.~Schuler {\it et al.}, \npb{523}, 423 (1998).
\bibitem{prevWork}
P.~del~Amo~Sanchez \etal\ (\babar\ Collaboration), \jprd{84}, 012004 (2011). 
\bibitem{Isgur} T. Barnes, S. Godfrey, and E.S. Swanson,  \jprd{72}, 054026 (2005).
\end{thebibliography}
